\documentclass[11pt,a4paper]{article}
\usepackage{bm}
\usepackage{amsmath}
\usepackage{graphicx}

\makeatletter

\usepackage{pos}

\title{Negative energy effects in processes involving bound particles: 
finite nuclear size effects}
 \ShortTitle{Negative energy, bound particles}

\author*[a]{Andrzej Czarnecki}

\affiliation[a]{Department of Physics, University of Alberta,\\ Edmonton,
  Alberta T6G 2E1, Canada}

\emailAdd{andrzejc@ualberta.ca}

\abstract{Probability of finding negative energy states in a hydrogen atom 
is decreased when the nucleus is extended. Reduced electron Compton wavelength is 
the characteristic size of the region around the nucleus
where negative energy contributions to the wave function are sizeable. }

\FullConference{%
  Loops and Legs in Quantum Field Theory - LL2022,\\
  25-30 April, 2022\\
  Ettal, Germany
}

\renewcommand{\logo}{\relax}
\renewcommand{\hookAfterAbstract}{%
  \par\bigskip
  \textsc{Alberta Thy 5-22}
}

\makeatother

\begin{document}
\maketitle

\global\long\def\calL{\mathcal{L}}%
\global\long\def\calM{\mathcal{M}}%
\global\long\def\calA{\mathcal{A}}%

\global\long\def\va{\bm{a}}%
\global\long\def\vb{\bm{b}}%
\global\long\def\vj{\bm{j}}%
\global\long\def\vk{\bm{k}}%
\global\long\def\vn{\bm{n}}%
\global\long\def\vp{\bm{p}}%
\global\long\def\vq{\bm{q}}%
\global\long\def\vr{\bm{r}}%
\global\long\def\vs{\bm{s}}%
\global\long\def\vu{\bm{u}}%
\global\long\def\vv{\bm{v}}%
\global\long\def\vw{\bm{w}}%
\global\long\def\vx{\bm{x}}%
\global\long\def\vy{\bm{y}}%
\global\long\def\vz{\bm{z}}%

\global\long\def\vA{\bm{A}}%
\global\long\def\vB{\bm{B}}%
\global\long\def\vD{\bm{D}}%
\global\long\def\vE{\bm{E}}%
\global\long\def\vF{\bm{F}}%
\global\long\def\vH{\bm{H}}%
\global\long\def\vJ{\bm{J}}%
\global\long\def\vK{\bm{K}}%
\global\long\def\vL{\bm{L}}%
\global\long\def\vN{\bm{N}}%
\global\long\def\vP{\bm{P}}%
\global\long\def\vR{\bm{R}}%
\global\long\def\vS{\bm{S}}%
\global\long\def\vU{\bm{U}}%
\global\long\def\vX{\bm{X}}%

\global\long\def\val{\bm{\alpha}}%
\global\long\def\vom{\bm{\omega}}%
\global\long\def\vga{\bm{\gamma}}%
\global\long\def\vep{\bm{\epsilon}}%
\global\long\def\vnabla{\bm{\nabla}}%
\global\long\def\vmu{\bm{\mu}}%
\global\long\def\vnu{\bm{\nu}}%
\global\long\def\vsi{\bm{\sigma}}%
\global\long\def\vSi{\bm{\Sigma}}%

\global\long\def\order#1{\mathcal{O}\left(#1\right)}%

\global\long\def\edge#1{\left.#1\right|}%
\global\long\def\d{\mbox{d}}%

\global\long\def\bra#1{\left\langle #1\right|}%
\global\long\def\ket#1{\left| #1 \right\rangle }%

\global\long\def\G{\widetilde{G}}%

\global\long\def\tr{\mbox{Tr}}%

\global\long\def\Li{\mbox{Li}_{2}}%

\global\long\def\az{\alpha_{Z}}%

\global\long\def\ap{\alpha_{\pi}}%

\global\long\def\za{Z\alpha}%

\global\long\def\Ep{E_{\bm{p}}}%

\global\long\def\Eb#1{E_{{\scriptscriptstyle \text{bind},#1}}}%

\section{Introduction}

The hydrogen atom consists most of the time of a nucleus and a single
electron. There is, however, a small probability that additional electron-positron
pairs appear. Positrons are represented by negative-energy solutions
of the Dirac equation. The probability of finding negative-energy
components of the wave function was first determined by Bethe \cite{Bethe:1948-H}.
In the ground state of a hydrogen-like ion with the atomic number
$Z$ that probability is
\begin{equation}
P_{-}\left(Z\right)=\frac{8\alpha_{Z}^{5}}{15\pi},\label{eq:Bethe}
\end{equation}
where $\az=Z\alpha$ and $\alpha\simeq1/137$ is the fine structure
constant. Eq.~\eqref{eq:Bethe} applies in the case of a point-like
nucleus. Here we demonstrate that in the case of an extended charge
distribution in the nucleus, this probability $P_{-}$ decreases.
This is interpreted as follows. Negative energy states appear mainly
in the region of a strong potential, in analogy with the Klein paradox
\cite{Klein29,Holstein:1998aa,Hansen_1981}. When the nucleus is extended,
the region of the strong potential is removed.

In the present work we focus on hydrogen-like ions containing spin-1/2
electrons. Ions with bound spin-0 particles have recently been considered
in \cite{Czarnecki:2022wsw}, in the limit of a point-like nucleus.
For an excellent review of a variety of bound-state phenomena, see
for example Ref.~\cite{Hoyer:2021adf}.

\section{Models of an extended nucleus}

\subsection{Spherical shell}

Instead of the whole charge $Ze$ concentrated in a point, we consider
a uniform surface charge distribution on a spherical shell of radius
$R$. The reason why we choose this particular model is that it is
easy to treat analytically. Besides, the details of the charge distribution
are not important for us. All we want to determine is how the probability
of finding negative energy states decreases in the absence of the
region of a very strong field. Inside a spherical shell there is no
field at all and we can change the field outside by changing the radius
$R$ of the shell.

We use such units that $\hbar=c=\epsilon_{0}=1$. Then the electrostatic
potential energy of an electron interacting with the shell is (\cite{chandrasekhar2003newton},
Chapter 15)
\begin{equation}
V_{\text{shell}}\left(r\right)=\begin{cases}
-\frac{\az}{R} & r<R,\\
-\frac{\az}{r} & r>R.
\end{cases}
\end{equation}
This function has a Fourier transform that is easy to remember: it
is similar to the Fourier transform of the Coulomb potential, $V_{\text{C}}\left(k\right)=-4\pi\az/k^{2}$,
but it is modulated by the sinc function,
\begin{equation}
V_{\text{shell}}\left(k\right)=V_{\text{C}}\left(k\right)\cdot\frac{\sin\kappa}{\kappa},\quad\kappa=kR.\label{eq:Vk}
\end{equation}
In the limit $R\to0$ we reproduce the Coulomb potential of a pointlike
nucleus. 

\subsection{Uniformly charged ball}

\begin{figure}

\centering\includegraphics[scale=0.9]{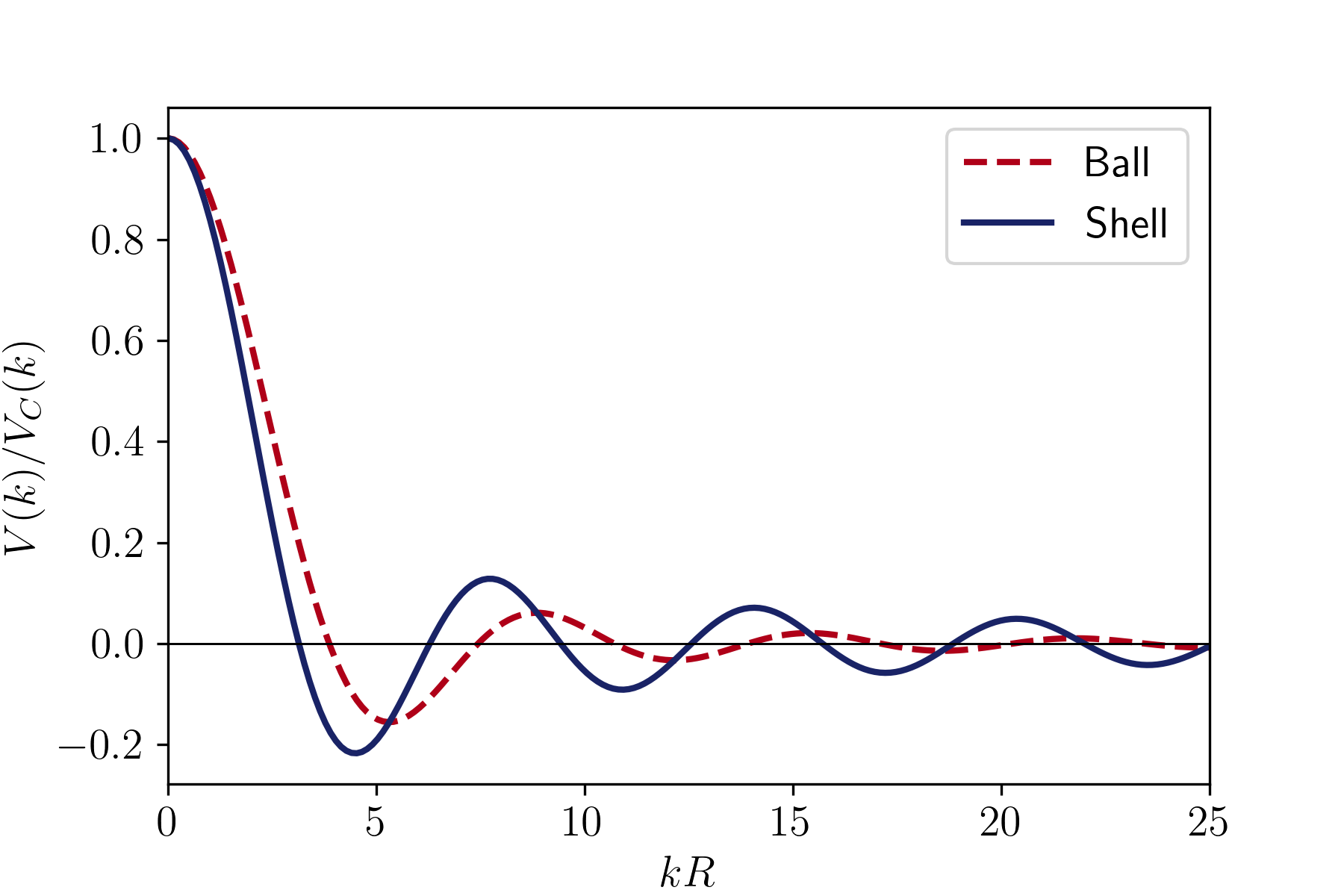}

\caption{Fourier transforms of electron's potential energy due to its interaction
with two charge distributions: a solid ball (red, dashed) and a hollow
shell (blue, solid), normalized to the Coulomb potential of a point-like
distribution. For a given radius of the charge distribution, the two
Fourier transforms are similar. In both cases, the smearing of the
charge decreases the magnitude of high-momentum (large $kR$) Fourier
components.\label{fig:Fourier}}

\end{figure}

Although we will be using the spherical shell model, here we introduce
a uniformly charged sphere as an alternative charge distribution,
\begin{equation}
\rho\left(r\right)=\begin{cases}
\text{constant} & r<R,\\
0 & r>R.
\end{cases}\label{eq:rhoBall}
\end{equation}
Our goal is to demonstrate that the Fourier spectrum of the potential
is similar in both cases, strengthening the argument that the details
of the charge distribution are not decisive for our conclusions. Electron's
potential energy resulting from the interaction with the density in
Eq. \eqref{eq:rhoBall} is
\begin{equation}
V_{\text{ball}}\left(r\right)=\begin{cases}
-\frac{\az}{R}\left(\frac{5}{4}-\frac{r^{4}}{4R^{4}}\right) & r<R,\\
-\frac{\az}{r} & r>R.
\end{cases}
\end{equation}
Its Fourier transform is
\begin{equation}
V_{\text{ball}}\left(k\right)=5V_{\text{C}}\left(k\right)\cdot\frac{3\left(\kappa^{2}-2\right)\sin\kappa-\kappa\left(\kappa^{2}-6\right)\cos\kappa}{\kappa^{5}},\quad\kappa=kR.\label{eq:Vballk}
\end{equation}
Fourier transforms of both the shell and the solid ball potentials,
normalized to the Coulomb potential, are shown in Figure \ref{fig:Fourier}.

\section{Negative energy states in the case of an extended nucleus}

In order to find the probability $P_{-}\left(Z\right)$ of finding
negative energy states in the potential of Eq. \eqref{eq:Vk}, we
need the negative energy component of the wave function $\phi_{-}\left(\vk\right)$.
Once that is determined, the needed probability is 
\begin{equation}
P_{-}\left(Z\right)=\int\frac{\d^{3}k}{\left(2\pi\right)^{3}}\left|\phi_{-}\left(\vk\right)\right|^{2}.
\end{equation}
That wave function component satisfies an integral equation, derived
from the Fourier transform of the Dirac equation \cite{Feshbach:1958wv}.
That equation is greatly simplified in the approximation where $\phi_{-}\left(\vk\right)$
is small and can be neglected under the integral \cite{Bethe:1948-H,Czarnecki:2022wsw}.
In the case of the hollow shell potential in Eq.~\eqref{eq:Vk} one
finds
\begin{equation}
\phi_{-}\left(\vk\right)=-\frac{4\pi\az}{\left(m+E\right)k^{2}}\frac{\sin kR}{kR}\sqrt{\frac{E-m}{2E}}\psi\left(0\right)
\end{equation}
where $E=\sqrt{m^{2}+k^{2}}$, $m$ is the electron mass, and the
wave function at the origin is approximately $\psi\left(0\right)\simeq1/\sqrt{\pi a^{3}}$,
where $a=1/\left(m\az\right)$ is the Bohr radius. The negative energy
probability can now be calculated,
\begin{equation}
P_{-}\left(Z\right)=\frac{4\az^{5}m^{3}}{\pi R^{2}}\int_{0}^{\infty}\d k\frac{\sin^{2}kR}{E\left(E+m\right)^{4}\left(E-m\right)}.
\end{equation}
Introduce a new variable $\epsilon$, $E=\epsilon m$, $k\d k=m^{2}\epsilon\d\epsilon$,
and denote $mR$ by $\mu$,
\begin{equation}
P_{-}\left(Z\right)=\frac{4\az^{5}}{\pi\mu^{2}}J,\qquad J=\int_{1}^{\infty}\d\epsilon\frac{\sin^{2}\left(\sqrt{\epsilon^{2}-1}\mu\right)}{\left(\epsilon+1\right)^{9/2}\left(\epsilon-1\right)^{3/2}},\qquad\mu=mR.\label{eq:numP}
\end{equation}
When the radius $R$ of the charge distribution is large in comparison
with the reduced Compton wavelength of the electron $1/m$, $\mu$
is large, the sine function in the numerator oscillates rapidly, and
its square could in some integrands be replaced by $1/2$. In our
present case, this does not work because in the region $\epsilon-1\ll1$
the square root factor in the argument of the sine compensates the
large value of $\mu$. Thus we first add and subtract a simpler function
that has a similar behavior for $\epsilon$ near 1,
\begin{equation}
J=\int_{1}^{\infty}\d\epsilon\left\{ \frac{\sin^{2}\left(\sqrt{\epsilon^{2}-1}\mu\right)}{\left(\epsilon+1\right)^{9/2}\left(\epsilon-1\right)^{3/2}}-\frac{\sin^{2}\left[\sqrt{2}\sqrt{\epsilon-1}\mu\right]}{2^{9/2}\left(\epsilon-1\right)^{3/2}}+\frac{\sin^{2}\left[\sqrt{2}\sqrt{\epsilon-1}\mu\right]}{2^{9/2}\left(\epsilon-1\right)^{3/2}}\right\} .\label{eq:J}
\end{equation}
In the first two terms we replace the sine-squared factors by 1/2;
this eliminates the dependence of these terms on $\mu$, they become
subleading for large $\mu$ and can be neglected. The last term, evaluated
analytically, gives
\begin{equation}
P_{-}\left(Z\right)\xrightarrow{\mu\gg1}\frac{\az^{5}}{4\mu}.\label{leadingLarge}
\end{equation}
We see that the probability of negative energy contributions decreases
with the radius of the charge distribution, confirming the intuition
that these contributions arise in the region of a strong potential. 

Subleading terms in Eq.~\eqref{eq:J} introduce a small negative
term. Including them significantly increases the range of $\mu$ over
which the large-$\mu$ asymptotics agrees with the exact $J$,
\begin{equation}
P_{-}\left(Z\right)\xrightarrow{\mu\gg1}\az^{5}\left(\frac{1}{4\mu}-\frac{16}{35\pi\mu^{2}}\right).\label{eq:largeR}
\end{equation}

\begin{figure}
\centering\includegraphics[scale=0.9]{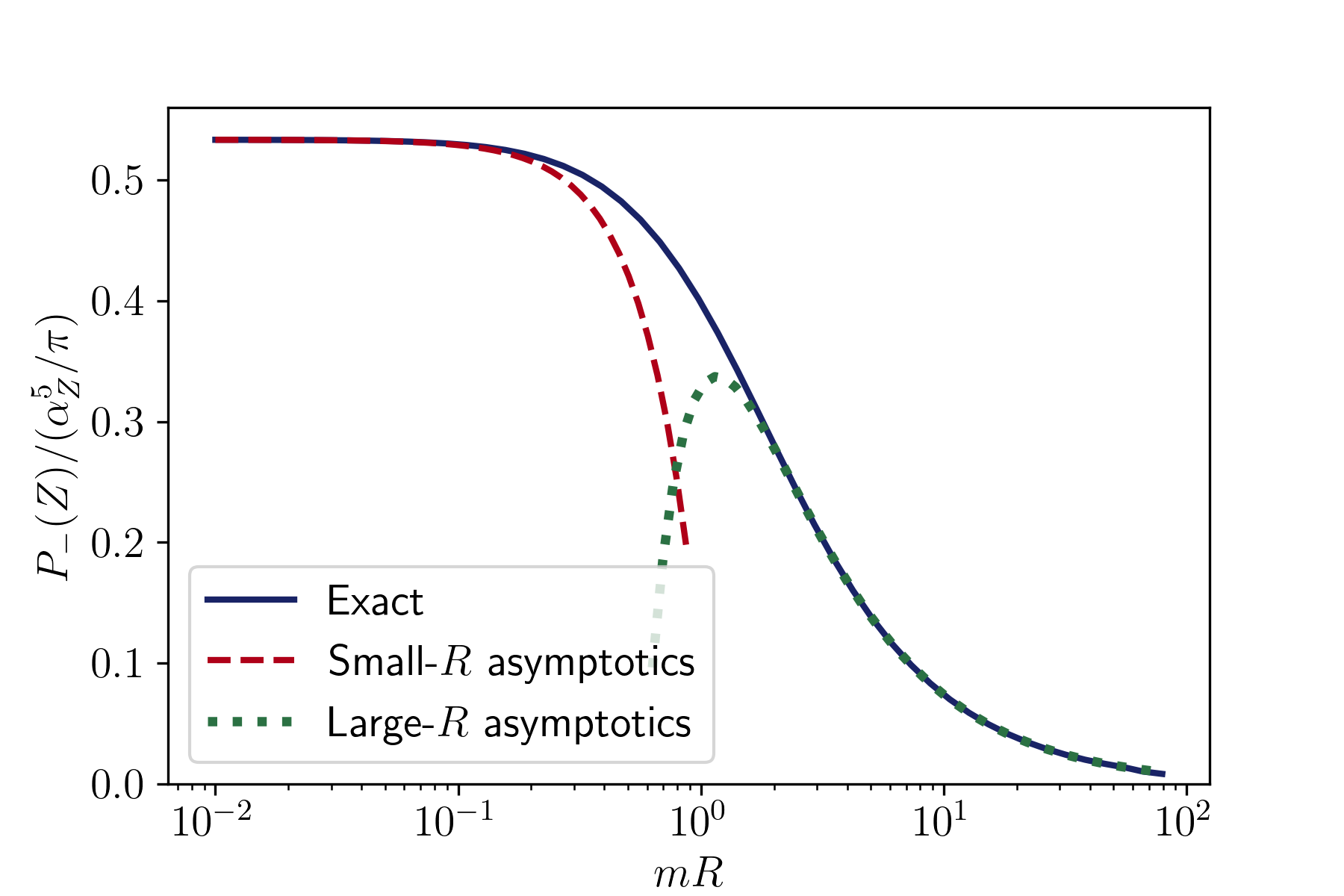}

\caption{Probability of finding negative energy states in an hydrogen-like
ion with an extended nucleus: a shell of radius $R$. The solid blue
curve shows the numerical integral as in Eq.~\eqref{eq:numP}. For
$R$ small in comparison with the electron's reduced Compton wavelength
(red, dashed line defined by Eq.~\eqref{eq:smallR}), the result
approaches Bethe's value of Eq.~\eqref{eq:Bethe}, with $8/15\simeq0.53$.
For large $R$ (greed, dotted line defined by Eq.~\eqref{leadingLarge}),
negative energy states are suppressed because the potential is weak.
\label{fig:Probability}}
\end{figure}

For completeness, we determine the behavior of $J$ for small $\mu$.
The sine function can be replaced by the first two terms of its Taylor
expansion. We retain only the square of the leading term and its product
with the subleading term and find
\begin{equation}
P_{-}\left(Z\right)\xrightarrow{\mu\ll1}\frac{4\az^{5}}{45\pi}\left(6-5\mu^{2}\right).\label{eq:smallR}
\end{equation}
When $\mu\to0$, this reproduces Bethe's result \cite{Bethe:1948-H}.
Both the large and small $R$ asymptotics of $P_{-}\left(Z\right)$
are plotted in Figure \ref{fig:Probability}.

\section{Conclusions}

Finite size of the nucleus removes the region where the potential
is very strong. This decreases the probability of finding negative
energy components in the electron wave function. For large radii of
the charge distribution the decrease is linear in the ratio of the
the electron's reduced Compton wavelength to the nuclear radius.

\section*{Acknowledgement}

I thank David Broadhurst for suggesting the point of view explored
in this work. This research was supported by Natural Sciences and
Engineering Research Canada (NSERC). 


\end{document}